\documentstyle[twocolumn,epsfig,aps]{revtex}

\begin{document}
\title{Nonadiabatic effects of atomic motion inside a high Q optical cavity}
\author{P. Zhang$^{1}$, Y. Li$^{1}$, C. P. Sun$^{1}$, and L. You$^{1,2,3}$}
\address{$^1$Institute of Theoretical Physics, The Chinese
Academy of Sciences, Beijing 100080, China}
\address{$^2$Interdisciplinary Center of Theoretical Studies,
The Chinese Academy of Sciences, Beijing 100080, China}
\address{$^3$School of Physics, Georgia Institute of Technology,
Atlanta, GA 30332, USA}
\date{\today }
\maketitle

\begin{abstract}
We revisit the topic of atomic center of mass motion of
a three level atom Raman coupled
strongly to an external laser field and the quantum field of a high Q
optical cavity. We focus on the motion related nonadiabatic effects
of the atomic internal dynamics and provide a quantitative
answer to the validity regime for the application of
the motional insensitive dark state as recently suggested
in Ref. [Phys. Rev. A {\bf 67}, 032305 (2003)].
\end{abstract}

\pacs{03.67.Lx, 89.70.+c, 32.80.-t}



\section{Introduction}

The development of quantum information science and technology carries the
potential of revolutionary impact on many aspects of our society, as
evidenced already by the applications in quantum cryptography, quantum
communication, and rudimentary quantum computing.
Among the physical systems being investigated,
high Q optical cavities coupled with trapped atomic qubits
represent a paradigm for this burgeoning field. In addition to their
demonstrated abilities for controlled (and coherent) quantum dynamics
of both atomic and/or cavity photonic qubits,
cavity QED systems are unique because they
represent a proto-type enabling technology for the coherent inter-conversion
of quantum information encoded in material qubits or flying photonic qubits
for propagation to far away places.

Despite much effort and spectacular advances from several groups in recent
years \cite{kimble,hood,cqed1}, high fidelity deterministic logic
operations even at the level of two qubits remain elusive in cavity QED
based systems. Among the factors as commonly attributed to being significant
road blocks, the localization of atomic motional wave packet is perhaps the
most demanding. In nearly all quantum computing protocols of atoms coupled
to a high Q cavity field, it is essential to reach the so-called strong
coupling limit, where the coherent coupling of an atom with the near
resonant cavity mode $g$ has to be much larger than both the cavity decay
rate $\kappa$ (one side) and the atomic spontaneous emission rate $\gamma$,
i.e. $g\gg\kappa$ and $g\gg\gamma$. Since $g^2$ is inversely proportional to
the mode volume of the cavity, it typically points to small cavities in the
Fabry-Perot arrangement, where the cavity mode is that of a standing wave
given by
\begin{eqnarray}
g(\vec r)&&=g_0\chi(\vec r),\nonumber\\
\chi(\vec r)&&={\frac{w_0}{w(z)}}
\exp\left[-{\frac{\rho^2}{w^2(z)}}\right]\sin(kz),
\end{eqnarray}
with $\rho=\sqrt{x^2+y^2}$ the transverse (polar) coordinate measured from
the cylindrically symmetric cavity axis along $z$ direction. The typical
geometries have the mode waist $w(z)=w_0\sqrt{1+z^2/z_0^2}$ much larger than
the cavity wavelength $\lambda$, where $z_0={\pi w_0^2/\lambda}$ is the
Rayleigh range. Unless each individual atomic motion is localized to much
less than the resonant wave length $\lambda$, i.e. in the so-called
Lamb-Dicke limit (LDL), this position dependent uncertainty of coupling
strength $g(\vec r)$ generally spoils the quantum coherence, and prevents
high fidelity quantum logic operations. This challenging limit is not so-far
under complete experimental control, it is especially problematic for
optical cavity QED systems.

\begin{figure}[tbp]
\hskip 36pt \includegraphics[width=1.75in]{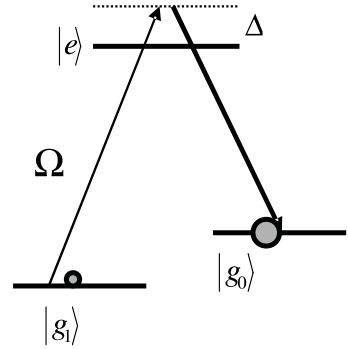}\\
\caption{Illustration of the Raman coupling.}
\label{fig1}
\end{figure}

Recently, two independent groups \cite{kennedy,duan} have noticed an
interesting scenario where the above undesirable position dependence of $g(%
\vec r)$ can be largely overcome with the use of a so-called dark state,
when the classical Raman laser field is assumed to have the same spatial
dependence as the quantum field $g(\vec r)$. For a large class of quantum
computing protocols based on atomic cavity QED,
the building block consists of a
three level $\Lambda$-type atom of stable ground states $|g_0\rangle$ and $%
|g_1\rangle$ that couple to an excited state $|e\rangle$. Such an
arrangement allows for a coherent mapping of an atomic qubit
$(\alpha|g_0\rangle+\beta|g_1\rangle)$ into the photonic coherence of the
cavity. In the most publicized version as originally suggested \cite{parkins}
, it is assumed that a classical laser field and the quantum cavity field
establishes a two-photon matched Raman resonance between $|g_1\rangle
\leftrightarrow|e\rangle$ and $|g_0\rangle\leftrightarrow|e\rangle$. By
denoting
the Rabi-frequency of the classical laser field as $\Omega$, the dipole
coupling in the interaction picture can be summarized as
\begin{eqnarray}
H_0 &=&-\hbar\Delta |e\rangle\!\langle e| \nonumber\\
&&+\hbar\Omega|e\rangle\!\langle g_1|
+\hbar g|e\rangle\!\langle g_0|c +h.c.,
\label{hm}
\end{eqnarray}
where $c$ is the annihilation operator for the near resonant
cavity photon mode used for Raman coupling, and the
common detuning $\Delta=\omega_L-\omega_{eg_1}=\omega-\omega_{eg_0}$.
It is easy to check that the following ``dark state"
\begin{eqnarray}
|D\rangle={\frac{1}{\sqrt{|g|^2+|\Omega|^2}}}\left(g
|g_1,0\rangle-\Omega|g_0,1\rangle\right),
\label{dk}
\end{eqnarray}
is an eigenstate of the Hamiltonian Eq. (\ref{hm}) with a zero eigenvalue,
where $|0\rangle_C$ and $|1\rangle_C$ denote the Fock state of $0$ or $1$
cavity photons respectively.
It is dark, or, immune to atomic spontaneous emission
because it contains no atomic excited state. By engineering a
counter-intuitive pulse sequence as in the STIRAP
(stimulated Raman adiabatic passage), and assume the
atom + cavity system to adiabatically follow the above dark state,
it leads to a highly efficient protocol for converting the atomic
qubit state into a photonic superposition according
to \cite{parkins,law,kuhn}
\begin{eqnarray}
(\alpha|g_0\rangle+\beta|g_1\rangle)\otimes|0\rangle_C\rightarrow
|g_1\rangle\otimes (\alpha|0\rangle+\beta|1\rangle)_C.
\end{eqnarray}

When atomic motion is considered, the position dependence of $g(\vec r)$
generally leads to a loss of coherence due to the potential
entanglement between the motion and the atomic internal
state as well as the cavity photon state. The
idea of the motional insensitive protocol \cite{kennedy,duan}, assumes a
classical laser field that has the same dependence as $g(\vec r)$. Following
the notation of Duan {\it et. al.} \cite{duan}, this assumption amounts to
\begin{eqnarray}
\Omega(\vec r,t)=\Omega_0(t)\chi(\vec r)=r_0g_0\alpha(t)\chi(\vec r).
\label{sm}
\end{eqnarray}
A simple arrangement involves choosing the pump and the cavity transitions
$|g_1\rangle\leftrightarrow|e\rangle$ and $|g_0\rangle
\leftrightarrow|e\rangle$ to correspond the left and right circular
polarized component of the same cavity mode. A more flexible setup would
involve the use of a different cavity mode such that near the cavity center,
$\chi(\vec r)$ for the two modes remain almost matched as the two modes
differ very little in their respective wavelengths \cite{ye}.

The aim of this paper is to study in detail nonadiabatic effects of the
above motion insensitive protocol. As was also noted in the Ref. \cite{duan}
, it clearly becomes difficult to maintain adiabaticity when the atomic
Raman coupling is too weak to affect the transfer, particularly near regions
of small $g(\vec r)$ values. Furthermore, an atom remaining in the dark state
essentially experiences no light force from the combined fields of both the
cavity mode and the external laser. This arguably leads to an upper limit on
the atomic kinetic energy; the duration $T$ for the STIRAP is determined by
atomic internal state dynamics. Thus for an atom with a velocity of $v_a$,
during the STIRAP, it will move a distance $v_aT$ if it is to remain in the
dark state. A larger $v_a$ leads simply to a large travelling distance.
Nonadiabatic effects will arise if atoms were to travel far enough to cross
nodal planes of $\chi(\vec r)$ (as we will see later this contradicts
the discussion of a better satisfied adiabatic condition near nodal points
as in Ref. \cite{duan}). The ideal operation of the motional
insensitive protocol would require the use of trapped atoms, i.e., with
atoms confined near regions of maximal $g(\vec r)$ \cite{duan} by an
external force independent of the cavity or Raman field. The trap has to be
strong enough to limit the atomic motion due to an amplitude smaller than
half the cavity wavelength $\lambda/2$, such that nodal crossing can be
completely avoided.

Another motivation for this study is the desire to understand nodal
crossing dynamics in general for the Raman configuration when
the trapping provided by an extra higher order cavity mode is absent
as in earlier experiments \cite{hood}, where either the resonant cavity
field or an external laser field provided confinement of atoms (not in the
dark state). Based on the dressed energy levels of an atom coupled to both
fields, we find that the dynamics of an atom crossing a nodal plane can be
effectively described in terms of
the celebrated Landau-Zener theory \cite{lz}. Yet, a
somewhat puzzling situation arises according to Landau-Zener theory which
prefers to have a larger velocity $v_a$ during the crossing in order to
maintain in the initial (dark) state. Consistent with the paper of
Duan {\it et. al.} \cite{duan}, the motional state insensitive protocol
works only in the limit when the atoms are trapped by yet an additional
mechanism such that its motion is limited to a variation of $g(\vec r)$
within approximately a factor of 2. On the other hand, a large velocity
tends to cause large amplitude motions, thus against the localization of the
Lamb Dicke limit. We thus find it interesting to study the relevant
Landau-Zener transitions in order to shed light on the motional
effects of atoms in cavity QED.

This paper is organized as follows. In section II, we formulate the
model of our study and illustrate parameter regimes of interests to current
experimental efforts. Sections III and IV are devoted respectively to
the study of the nonadiabatic level crossing in terms of a Landau-Zener
transition dynamics and the comparison between numerical simulations
and the approximate analytic Landau-Zener state transition formulae.
We have developed an interesting analytic mapping
(in the absence of an external trap) of the atomic
motion through a nodal point into a Landau-Zener level crossing dynamics.
Within each of the above sections, we will study various limiting cases,
mainly focusing on a simple model that involves a 1-dimensional
motion along the cavity axis \cite{ad,you}. Finally we
summarize and attempt to make some general conclusions in Sect. V. The
appendices contain several technical points that may be
useful for related studies.

\section{formulation}
In the descriptions to follow, we will assume both fields to be on resonance
and take the atomic detuning $\Delta=0$. For the more general situation as
shown in the appendix \ref{apda} with a nonzero but constant
$\Delta\neq 0$ (position independent), we find it simply leads to
formally identical results as discussed here for $\Delta=0$ (see appendix
\ref{apda}).
When necessary, an
external trap, assumed to be internal state independent is assumed to be
available and centered around the maximum of $g(\vec r)$ to confine
atomic motion to within the order of half the standing wave wavelength
as in Ref. \cite{duan}.

The Hamiltonian for atomic internal degrees of freedom
(on resonance $\Delta=0$) can be written as a
$3\otimes3$ matrix \cite{duan} in the basis $\{|e,0\rangle,|g_0,1
\rangle,|g_1,0\rangle\}$,
\begin{eqnarray}
H=\hbar\left({\
\begin{array}{ccc}
0 & g & \Omega \\
g^* & 0 & 0 \\
\Omega^* & 0 & 0
\end{array}
}\right).
\end{eqnarray}
Within the semiclassical approximation for the
atomic motion, the center of mass motion is described by $\vec p^2/2M$ with $M$ the atomic
mass and $\vec p$ the atomic momentum. While for slow atoms, we may wish to
include the associated forces due to the atomic interaction with the
spatially dependent laser (cavity) fields. We will first of all, consider
the simple case of predetermined atomic motion as corresponds to atoms
staying in the dark state, i.e. we simply assume that atomic motion is not
affected. We will also comparatively address the
case of a trapped harmonic atom motion. We defer the
inclusion of atomic dipole forces to a future investigation.
With these assumptions,
we find the three eigenvalues
\begin{eqnarray}
E_0&=&0,  \nonumber \\
E_\pm&=&\pm\chi[r(t)]g_0\sqrt{1+|r_0\alpha(t)|^2},
\label{edress}
\end{eqnarray}
with $E_\pm$ depends on $\vec r$.
Here, we have used $\Omega=r_0\alpha(t)g$,
$|g|=|\chi[r(t)]g_0|$, and $\sqrt{|g|^2+|\Omega|^2}=|\chi[r(t)]g_0|
\sqrt{1+|r_0\alpha(t)|^2}$.
We note that $E_{+}>E_{-}$ when $\chi[r(t)]g_0>0$, and
$E_{+}<E_{-}$ when $\chi[r(t)]g_0<0$.

\begin{figure}[tbp]
\includegraphics[width=3.in]{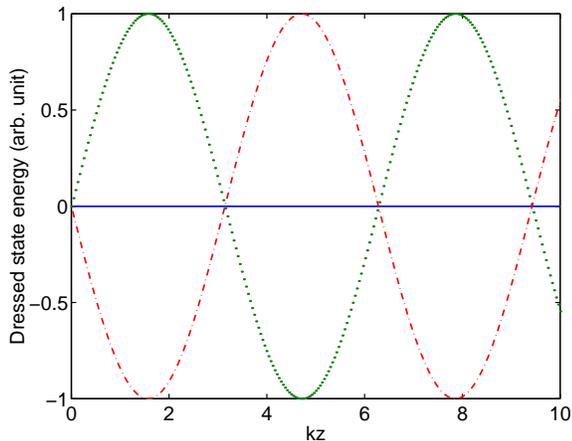}
\caption{Dressed energies of the coupled atom + cavity system
at a particular radial location along the cavity axis $z$
(at a fixed time $t$).
When the atom moves through a nodal plane, nonadiabatic
transfer of its internal state out of the dark state
becomes a critical issue, which constitutes the main topic being
studied in this work.}
\label{fig2}
\end{figure}

Figure \ref{fig2} shows the atomic dressed state energies (\ref{edress})
along the cavity axis at a given radial location ($\Delta=0$)
for a fixed time $t$ when the external laser is tuned on. For an
atom to cross a nodal plane, it has to be initially within a distance
reachable within the duration of the external pump pulse $\alpha(t)$
with its initial velocity.

In addition to dark state (\ref{dk}), the two other
eigenstates are
\begin{eqnarray}
|B_\pm\rangle &=&{\frac{1}{\sqrt 2}}(|B\rangle\pm|e,0\rangle),  \nonumber\\
|B\rangle &=&\frac{e^{i\phi}}{\sqrt{1+|r_0\alpha(t)|^2}}[|g_0,1\rangle
+r_0^*\alpha(t)|g_1,0\rangle],
\label{bk}
\end{eqnarray}
where $e^{i\phi}=g^*/|g|$ and will assumed to be unity later.

Before writing down the Schrodinger equation, we absorb the phase factors
due to the adiabatic evolution along each of the above (time-dependent)
eigenstate
\begin{eqnarray}
|\Psi(t)\rangle =\sum_{n=0,+,-}C_n(t)e^{-i\int_0^tE_n(t)dt'/\hbar}|n\rangle,
\label{eps}
\end{eqnarray}
where we have used the shorthand notation of $n=0$ for the dark state
$|D\rangle$ and $n=\pm$ for states $|B_\pm\rangle$ respectively.
The coefficients $C_n$ are governed by
\begin{eqnarray}
\dot{C_0} &=&-e^{-i\int_0^tE_{+}(t')dt'/\hbar}\langle D|%
\dot B_+\rangle C_{+}  \nonumber \\
&& -e^{-i\int_0^tE_{-}(t')dt'/\hbar}\langle D|\dot B_-\rangle C_{-},  \nonumber \\
\dot C_+ &=&-e^{+i\int_0^tE_{+}(t')dt'/\hbar}\langle B_+|%
\dot{D}\rangle C_0  \nonumber \\
&& -e^{-i\int_0^t [E_{-}(t')-E_{+}(t')] dt'/\hbar}\langle B_+|\dot B_-\rangle C_{-},  \nonumber \\
\dot C_- &=&-e^{i\int_0^tE_{-}(t')dt'/\hbar}\langle B_{-}|%
\dot{D}\rangle C_0  \nonumber \\
&& -e^{-i\int_0^t [E_{+}(t')-E_{-}(t')] dt'/\hbar}\langle
B_{-}|\dot B_+\rangle C_{+}.
\label{oeqs}
\end{eqnarray}

We now assume that $\alpha(t)$ is a real parameter, which leads to
\begin{eqnarray}
\langle B_\pm|\dot{B}_\mp\rangle &&=\frac{1}{2}\langle B|\dot{B}\rangle =0,
\nonumber \\
\langle D|\dot{B_\pm}\rangle &&=\frac{1}{\sqrt{2}}\langle D|\dot{B}\rangle=%
\frac{r_0^*\dot{\alpha}(t)}{\sqrt{2}(1+|r_0\alpha(t)|^2)}.
\end{eqnarray}

Denoting $E_\pm=\pm \varepsilon$, we end up with the simplified form of Eqs.
(\ref{oeqs})
\begin{eqnarray}
\dot{C_0} &=&-e^{-i\int_0^t\varepsilon (t')dt'/\hbar}K(t)
C_{+}-e^{i\int_0^t\varepsilon (t')dt'/\hbar}K(t) C_{-},
\nonumber \\
\dot C_+ &=&e^{+i\int_0^t\varepsilon (t')dt'/\hbar}K^*(t)
C_0,  \nonumber \\
\dot C_- &=&e^{-i\int_0^t\varepsilon (t')dt'/\hbar}K^*(t)C_0,
\label{eqd}
\end{eqnarray}
with
\begin{eqnarray}
K(t)&=&\frac{r_0^*\dot{\alpha}(t)}
{\sqrt{2}(1+|r_0\alpha(t)|^2)}\nonumber\\
&=&\langle D|\dot{B}_\pm\rangle.
\label{eq14}
\end{eqnarray}

These equations can be numerically integrated using standard algorithms to
investigate nonadiabatic level crossings. Before attempting an analytical
understanding of the level crossing dynamics near the nodes of the cavity
mode function $\chi(\vec r)$ in the next section, we first consider here
typical regimes of system parameters.

\subsection{parameters}

We will use Cs as a proto-type atom for the estimation of various atomic
parameters. The resonant transition between
the 6$^2S_{1/2}$ and 6$^2P_{3/2}$,
(unclear spin $I=7/2$), occurs at around $\lambda=852.35$ (nm),
$\hbar\omega=1.455$ (eV) and excited state lifetime $(1/\gamma)=30.70$ (ns),
or $\gamma=(2\pi)5.18$ (MHz). The recoil frequency of resonant transition is
about $\omega_R=\hbar k^2/(2M)=2.07$ (kHz), corresponding to a temperature
of 0.198 ($\mu$K), or a recoil velocity of $0.352$ (cm/s). The wave length
of $\lambda_{{\rm FORT}}=936$ (nm) used for the dipole trapping of atoms in
Ref. \cite{duan} corresponds to a higher order cavity field.

The vacuum Rabi coupling between the cavity field and the atom is
$g_0=(2\pi)50$ (MHz) as in the recent CalTech experiment \cite{kimble}.
For order of magnitude estimates, the ratio $r_0={\Omega}/{[g\alpha(t)]}$
can be taken as unity.

According to Sect. III. (A) of Ref. \cite{duan}, we take the adiabatic
parameter $\alpha(t)$ as a Gaussian function
\begin{equation}
\alpha (t)=\alpha_0\exp\left[-{(t-T_0)^2\over t_W^2}\right],
\label{at}
\end{equation}
with the total protocol for state transfer being
approximated as $T=2T_0$. The amplitude $\alpha_0$
is assumed to be $30$ and the width $t_W$ is assumed to be $T_0/3$.
For the efficient operation of the quantum state transfer protocol
utilizing the dark state adiabatic passage, the excited state atomic
life time $\propto 1/\gamma$ essentially sets the time scale,
for this reason we take $t_W\sim 1/\gamma$, or $T_0\sim 10/\gamma$.

Assuming that atoms are trapped inside a single well of the standing wave
cavity field, or of the wells of the dressed energy $E_\pm(\vec r)$, we can
estimate the oscillation frequency inside according to
\begin{eqnarray}
g_0\chi(\rho,z-\lambda/4)&&=g_0\exp\left[-{\frac{\rho^2}{w^2(z-\lambda/4)}}%
\right]\sin(kz-\pi/2)  \nonumber \\
&&=g_0\exp\left[-{\frac{\rho^2}{w^2(z-\lambda/4)}}\right][2\sin^2({\frac{kz}{%
2}})-1]  \nonumber \\
&&\approx g_0\exp\left[-{\frac{\rho^2}{w^2(z)}}\right][{\frac{(kz)^2}{2}}-1],
\end{eqnarray}
near the axis center where the well is deepest, which gives the strongest
axial oscillation and radial oscillation according to
\begin{eqnarray}
{\frac{1}{2}}M\omega_z^2 z^2&&={\frac{1}{2}} \hbar g_0k^2z^2, \\
{\frac{1}{2}}M\omega_\rho^2 \rho^2&&=\hbar g_0{\frac{\rho^2}{w^2(z)}},
\end{eqnarray}
i.e. we obtain
\begin{eqnarray}
\omega_z &&= \sqrt{2 g_0 (\hbar k^2/2M)},  \nonumber \\
\omega_\rho &&=2 \sqrt{g_0 (\hbar k^2/2M)}\,{\frac{1}{k w(z)}}.
\end{eqnarray}
Given the additional enhancement due to $\Omega(\vec r)$ in $E_\pm$, and use
the estimated parameters as outlined above, we then take $\omega_z\sim
(2\pi)500$ (kHz), and assume a fundamental cavity mode waist of $%
w(0)=30\lambda$, we end up with $\omega_\rho\sim (2\pi)2.65$ (kHz),
of the same orders of magnitude as in Ref. \cite{kimble}
of the optical trap from a higher
order cavity mode. Of course, these estimates are valid only
for atomic motion near the bottom of the trap. For
significantly higher atomic energies,
as for instance in the recent experiment \cite
{kimble}, where atomic kinetic energy is of the order of $(2\pi) 20$ (MHz),
or about half of the actual potential barriers at about $(2\pi) 50$ (MHz),
and about $40\,(\hbar\omega_z)$ and $7500\,(\hbar\omega_\rho)$,
simple harmonic motion cannot be assumed.

Within the regimes of these parameters, atomic center of mass motion is well
approximated by classical dynamics in a conservative potential. We can also
estimate the maximal velocity of these atoms when trapped in the above
single well potential,
\begin{eqnarray}
{\frac{1}{2}}Mv_{zM}^2(\hbar k/M)^2&&=40 (\hbar\omega_z)  \nonumber \\
&&\sim {\frac{1}{2}} Mv_{\rho M}^2(\hbar k/M)^2=7500 (\hbar\omega_\rho),
\end{eqnarray}
where we have expressed velocities in atomic recoil units.
Thus we find
\begin{eqnarray}
v_{zM}\sim v_{\rho M}&&
=\sqrt{\frac{40 ({\rm MHz})}{2\omega_R}} =\sqrt{10000}\nonumber \\
&&=100 (v_R)\approx 35\ ({\rm cm/s}).
\end{eqnarray}

This leads to the assumption of atomic velocities with the following choices
for numerical simulations; $v_z\approx 1$ (m/s) $\sim 0.036\, (\lambda\gamma)$
(10 times more kinetic
energy), $v_z\approx 0.35$ (m/s), and $v_z\approx 0.1$ (m/s) (10 times less
kinetic energy). With $T_0\sim 10/\gamma$, the respective distances a
typical atom travels during the state transfer protocols become $0.03-0.3$ ($%
\mu$m), which is a significant fraction of $\lambda/4$, half the distance
between the nearest nodal planes. It is important to emphasize that in this
limit which corresponds to the recent experiment \cite{kimble},
atomic kinetic energy is much higher than its single photon recoil energy,
and the atom's motional
quantum state is much higher than the ground state of each trapped well.
This lends strong support to our assumption of using a constant (predetermined)
atomic trajectory in studying the level crossing dynamics. When an
additional cavity field is used to confine atoms, optical dipole force
is the reason for atoms to turn around at classical turning points.

\subsection{Qualitative picture of the failure of adiabaticity}
\label{subsecb}
To maintain adiabaticity during atomic motion, the system must
satisfy \cite{cpsun}
\begin{eqnarray}
\left|\frac{\langle n(t)|\partial_{t}|m(t)\rangle}{E_n(t)-E_m(t)}\right|
\ll 1,
\label{adc}
\end{eqnarray}
for all adiabatic energy levels. The above notation applies to a general
time dependent Hamiltonian $H(t)$ with eigenfunctions $|m(t)\rangle$ and
eigenvalues $E_m(t)$. The time derivative of $|m(t)\rangle $ should be
calculated according to
\begin{eqnarray}
\partial_{t}|m(t)\rangle=\sum_{R_{i}}\frac{\partial}{\partial R_{i}}
|m(t)\rangle\times\frac{dR_{i}}{dt},
\end{eqnarray}
with $R_{i}$ the $i$th parameter.

In the problem considered here, there are four parameters:
$\alpha$, $x$, $y$, and $z$. Similar to the Ref. \cite{duan},
we take the spatial mode functions to be identical as in Eq. (\ref{sm}).
It is easy to see that the
dark state Eq. (\ref{dk}) $|D\rangle$ as well as the other two eigenstates
\begin{eqnarray}
|B_\pm\rangle=\frac{1}{2\sqrt{g^2+\Omega^2}}(g|g_0,1\rangle+\Omega
|g_1,0\rangle) \pm\frac{1}{\sqrt{2}}|e,0\rangle,
\end{eqnarray}
only depends on the parameter $\alpha$ and has nothing to do with $\vec r$.
So the numerator of Eq. (\ref{adc}) is independent of the atomic speed $%
\dot {\vec r}$. On the other hand, two of the three eigenvalues
\begin{eqnarray}
E_{\pm}&&=\pm\sqrt{g^2+\Omega^2},
\end{eqnarray}
do depend on $\vec r$. Thus the denominator of Eq. (\ref{adc}) depends on
the position of the atom $\vec r$. Assuming that the time evolution of $%
\alpha(t)$ is uncorrelated with atomic motion $\vec r(t)$, the value of
Eq. (\ref{adc}) may become large, especially in the regions where $
|E_n(t)-E_m(t)|\ll |\langle n(t)|\partial_{t}|m(t)\rangle|$.

Two regions require special attention: (1) the nodal planes
perpendicular to the cavity axis due to the standing wave term $\sin(kz)=0$;
and (2) the region away from the cavity axis due to the exponentially
damped Gaussian term $\exp[-\rho^2/\omega^2(z)]$.
In this study, we focus on the level crossing dynamics, which
happens mainly in the first region defined above.
For the latest
experiments, as in \cite{kimble,hood} where atoms are localized to a single
well along the cavity axis, we may expect reduced nonadiabaticity because no
actual level crossing occurs.

\begin{figure}[h]
\includegraphics[width=3in]{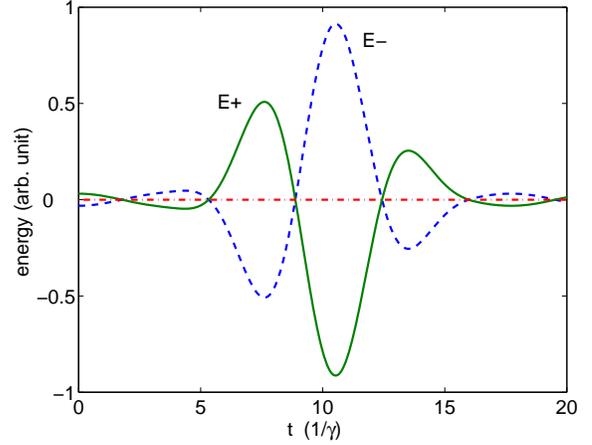}\\
\caption{The time dependent dressed state energies $E_{0/{\pm}}$
with the time dependent pulse shape $\alpha(t)$ as in Eq. (\ref{at}).
We have used $z(t)=vt$ with
the speed of the atom being $v=0.14(\lambda\gamma)\approx 4 $ (m/s).
$\alpha(t)$ is assumed to begin increasing at time $t=0$ when
the atom is located at the peak of the cavity field where $|\sin(kz)|=1$.}
\label{fig3}
\end{figure}

Before attempting a comprehensive understanding of the problem,
we will first discuss the qualitative picture of how adiabatic
following is violated during the atomic motion in this subsection.
For simplicity, we will model the atomic motion as being
simple one dimensional. From
Eq. (\ref{eq14}), the adiabatic condition Eq. (\ref{adc}) is just
\begin{eqnarray}
\left|\frac{K(t)}{E_{\pm}(t)}\right|\ll 1,
\label{ddd}
\end{eqnarray}
where $|E_{\pm}(t)|=|\chi\lbrack r(t)]|g_0\sqrt{1+|r_0\alpha (t)|^2}$ as
defined before. We see that $E_{\pm}(t)$ becomes sufficiently small
(see Fig. \ref{fig3}) when the atom is near the nodal plane $\sin(kz)=0$,
where the condition Eq. (\ref{ddd}) is easily violated.

After elementary substitutions, we find
\begin{eqnarray}
\left|\frac{K(t)}{E_{\pm}(t)}\right|=\frac{f(t)}{|\sin(kz)|},
\label{eee}
\end{eqnarray}
with
\begin{eqnarray}
f\left(t\right)=\left|\frac{\dot{\alpha}(t)}
{\sqrt{2}g_0(1+|\alpha(t)|^2)^{{3}/{2}}}\right|.
\label{ft}
\end{eqnarray}

\begin{figure}[h]
\includegraphics[width=3in]{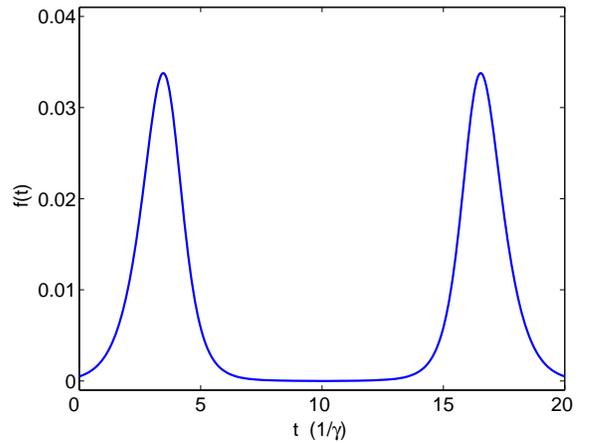}\\
\caption{The function $f(t)$ of Eq. (\ref{ft}).}
\label{fig4}
\end{figure}

In Fig. \ref{fig4}, we have graphed the function $f(t)$.

The two prominent features as in Fig. \ref{fig4} at times of
$3.3/\gamma$ and $16.7/\gamma$ correspond to the instants
when the Gaussian shaped pulse gives rise to largest shape changes.
Apparently, $f(t)$ is significant at $t=3.3/\gamma$ and
$t=16.7/\gamma$. From Eq. (\ref{eee}), we note
if the atomic position is near a nodal plane
when $|\sin(kz)|$ is small, $|\frac{K(t)}{E_{\pm}(t)}|$
becomes large and the adiabatic condition
(\ref{ddd}) can become severely violated at these instants.
This adiabatic breakdown is less serious at $t=16.7/\gamma$
when the internal state transfer protocol is almost completed,
but detrimental at $t=3.3/\gamma$, near the beginning
of the process.

\section{Landau-Zener transitions}
\label{sect3}
As was shown from the previous discussions, the application of
the adiabatic approximation can potentially fail near a nodal
plane where $\chi(\vec r)=0$. In this section, we hope to analytically
investigate the transition from the dark state to bright states when the
atom goes across a nodal plane.

\subsection{One-dimensional motion along the cavity axis}

We first deal with the one dimension case of atomic motion along
the cavity axis. Assuming
\begin{eqnarray}
\chi(\vec r)=\sin(kz),
\end{eqnarray}
and take the atomic motion according to $z=vt$, our problem is to solve Eq.
(\ref{eqd}), with the initial conditions
\begin{eqnarray}
\left| C_0(-u)\right| &=&1,  \nonumber \\
C_{\pm}(-u) &=&0.
\end{eqnarray}
More specifically, we in fact only wish to solve for $|C_0(u)|^2$
and $|C_{\pm}(u)|^2$ after one nodal crossing.

As the lowest order approximation, we assume $\alpha(t)$ and
$K(t)$ to be constants in the domain $kz\in [-u,u]$ and
we also assume that the energy $\varepsilon (t)/\hbar=\sin(
kz)g_0\sqrt{1+|\alpha(t)|^2}$ can be approximated as a linear
function of time (see appendix \ref{apdb}) in this domain
\begin{eqnarray}
\varepsilon (t)/\hbar \approx kvtg_0\sqrt{1+|\alpha(t)|^2},
\end{eqnarray}
where $v$ is a constant atomic speed along the cavity axis.

The second equation of Eq. (\ref{eqd}) can be rewritten as
\begin{eqnarray}
{\frac{d^2}{dt^2}C}{_{+}} &=&i\epsilon (t)\dot{C}_{+}/\hbar+e^{i\int_0^t%
\varepsilon (t')dt'/\hbar}K\dot{C_0}  \nonumber \\
&=&i\epsilon (t)\dot{C}_{+}/\hbar-e^{i\int_0^t\varepsilon
(t')dt'/\hbar}K  \nonumber \\
&&(e^{-i\int_0^t\varepsilon (t')dt'/\hbar}KC_{+}
+e^{i\int_0^t\varepsilon (t')dt'/\hbar}KC_{-})\nonumber \\
&=&i\epsilon (t)\dot{C}_{+}/\hbar-K^2C_{+}-e^{2i\int_0^t\varepsilon
(t')dt'/\hbar}K^2C_{-}.
\label{ggg}
\end{eqnarray}
Using
\begin{eqnarray}
\dot{C}_{-} &=&e^{-i\int_0^t\varepsilon (t')dt'/\hbar}KC_0
\nonumber \\
&=&e^{-2i\int_0^t\varepsilon (t')dt'/\hbar}\dot{C}_{+},
\end{eqnarray}
and $C_{\pm}(-u)=0$, we find
\begin{eqnarray}
C_{-} &=&\int_{-u}^te^{-2i\int_0^{t'}\varepsilon (t'')dt''/\hbar}
\dot{C}_{+}\left( t'\right)dt'  \nonumber \\
&=&e^{-2i\int_0^{t'}\varepsilon (t'')dt''/\hbar}C_{+}  \nonumber \\
&&+2i\int_{-u}^te^{-2i\int_0^{t'}\varepsilon (t'')dt''/\hbar}
{\varepsilon (t')}C_{+}(t')dt'/{\hbar}. \label{fff}
\end{eqnarray}
Substituting Eq. (\ref{fff}) to (\ref{ggg}), we find
\begin{eqnarray}
{\frac{d^2}{dt^2}}{C_{+}} &=&i{\epsilon (t)}\dot{C}/{\hbar}-2K^2C_{+}
\nonumber \\
&&+2iK^2\int_{-u}^te^{2i\int_{t'}^t\varepsilon (t'')dt''/\hbar}
{\varepsilon (t')}C_{+}(t')dt'/{\hbar}.
\label{sterm}
\end{eqnarray}
The last term is rapidly oscillating and within the lowest order
approximation, it can be neglected (see appendix \ref{apdc}). Then the
equations of $C_{+}$ and $C_0$ become
\begin{eqnarray}
{\frac{d^2}{dt^2}}{C_{+}}=i{\epsilon (t)}\dot{C}_{+}/{\hbar}-2K^2C_{+},
\label{hhh}
\end{eqnarray}
and
\begin{eqnarray}
\dot{C}_+=e^{i\int_0^t\varepsilon (t')dt'/\hbar}KC_0.
\end{eqnarray}

Using the transformation
\begin{eqnarray}
d_0 &=&(-i)\frac{1}{\sqrt{2}}C_0,  \nonumber \\
c_{+} &=&e^{-i\int_0^t\varepsilon (t')dt'/\hbar}C_{+},
\label{j}
\end{eqnarray}
we find that Eq. (\ref{hhh}) is equivalent to
a problem described by an effective Hamiltonian
as below
\begin{eqnarray}
i\hbar{\partial\over \partial t}
\left({\begin{array}{c}
d_0 \\
c_+
\end{array}}\right)=\left({\begin{array}{cc}
0 & -\sqrt{2}\hbar K \\
-\sqrt{2}\hbar K & \varepsilon
\end{array}}\right)\left({\begin{array}{c}
d_0 \\
c_+
\end{array}}\right),
\end{eqnarray}
with the initial condition
\begin{eqnarray}
d_0(-u) &=&(-i)\frac{1}{\sqrt{2}},  \nonumber \\
c_{+}(-u) &=&0.
\label{k}
\end{eqnarray}
Nonadiabatic effect
induced transitions mainly occur within the domain $kz\in[-u,u]$,
i.e. $u$ is chosen such that beyond this domain,
adiabatic condition Eq. (\ref{adc}) is well satisfied.
In the end, as we will see later that our result is independent
of the choice of $u$.
Since $\epsilon (t)$ is a linear function of $t$, this problem is
exactly the same one as discussed in the original Zener's paper
\cite{lz}. At the edge of the domain $kz\in[-u,u]$,
$|K|\ll |\epsilon|$ and
the initial conditions can be adiabatically maintained to the domain $%
(-\infty ,\infty )$, just as Zener has done. Noting the
normalization condition
\begin{eqnarray}
|d_0|^2+|c_0|^2=\frac{1}{2},
\end{eqnarray}
and using Zener's solution \cite{lz}, we find that
\begin{eqnarray}
|d_0(\infty)|^2=\frac{1}{2}\exp \left(-2\pi\frac{2K^2}{|kvg_0\sqrt{%
1+|\alpha|^2}|}\right) ,
\end{eqnarray}
which leads to
\begin{eqnarray}
|C_{+}(\infty)|^2&=&\frac{1}{2}-|d_0(\infty)|^2 \nonumber \\
&=&\frac{1}{2}\left[1-\exp\left(-2\pi\frac{h(t)}{|kvg_0|}\right)\right],
\label{m}
\end{eqnarray}
with
\begin{eqnarray}
h(t)=\frac{2K^2}{\sqrt{1+|\alpha|^2}}=\frac{\dot\alpha^2(t)}
{(\sqrt{1+|\alpha|^2})^{5/2}}.
\label{eqh}
\end{eqnarray}

This constitutes the main result of our paper.
We note the dark state probability after crossing a node becomes
\begin{eqnarray}
P&=&1-|C_{+}(\infty)|^2-|C_{-}(\infty)|^2\nonumber \\
&=&\exp\left(-2\pi\frac{h(t)}{|kvg_0|}\right).
\label{lza}
\end{eqnarray}
This leads to the conclusion that the larger the atomic speed $v_z$ is,
the
more reasonable it is to adopt the approximation of taking $\alpha(t)$
and $K(t)$ as constants and neglect the term $2iK^2\int_{-u}^t
\exp[{2i\int_{t'}^t\varepsilon (t'')dt''/\hbar}]{%
\varepsilon(t')}C_{+}(t')dt'/{\hbar}$. Furthermore as we shall
see in the next section, even in the limit of a small $v_z$, the
transition probability (\ref{lza}) as given by the analytic
Landau-Zener method also compares well with results from numerical
simulations.

\subsection{3-dimensional motion of atoms}

In the limit as considered presently when atomic motion is predetermined,
a full 3-dimensional center of mass motion of the atom can be discussed
without much further complications. Essentially, it is the component of
the atomic velocity along the cavity axis direction that is involved in the
level crossing dynamics, motion in the orthogonal directions only
causes the crossing to be at different radial locations, thus different
level spacing characteristics.

\section{results and discussions}

In this section, we investigate the dark state survival probability by
comparing the analytic result (\ref{lza}) under the Landau-Zener
approximation with numerical solutions of Eq. (\ref{eqd}).

Unless otherwise noted, the parameter $g_0=(2\pi)50$ (MHz) is used,
corresponds to the $\overline{g}$
at the end of Sect. II of Ref. \cite{duan}.
The atomic mass for Cs is $M\approx 133\times 1.67\times 10^{-27}$ (kg).

We assume the pump shape is given by Eq. (\ref{at}) where $\alpha_0=30$,
$T_0=3\times 10^{-7}$ (s) $\sim {10\gamma}$ (so that the total operation
occurs within $2T_0=20/{\gamma}$, as typical for the optimal STIRAP
process), and $t_W=T_0/3$. Therefore $t\in (0,2T_0)$.

\begin{figure}[h]
\includegraphics[width=3in]{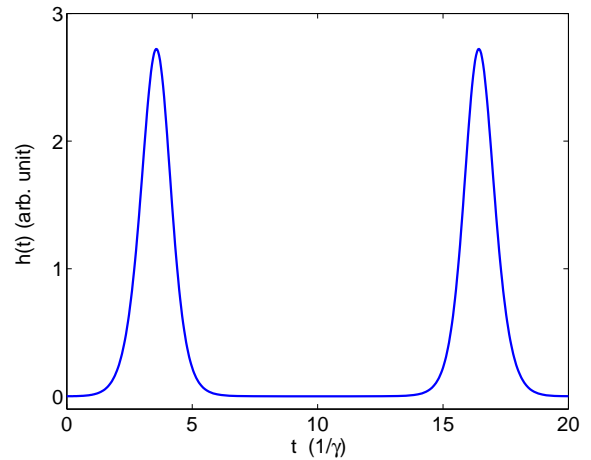}
\caption{The function $h(t)$.}
\label{fig5}
\end{figure}

From the approximate Landau-Zener result Eq. (\ref{lza}), it is
easy to see that the atom's dark state survival probability $P$
after crossing a node is
closely related to the function $h(t)$ of Eq. (\ref{eqh}).
The larger is $h(t)$, the smaller is $P$.
We show the time dependence of $h(t)$ in Fig. \ref{fig5}.
It resembles the function $f(t)$ as shown before in Fig. \ref{fig4}.
The duration of the state transfer protocol is taken
to be $20/\gamma$ so that the
(unavoidable) maximums of $h(t)$ are clearly displayed.

It is easy to see that at times near $t=3.3/\gamma$
or $t=16.7/\gamma$, $h(t)$ becomes rather large.
If an atom crosses a node at these instants,
the transition probability to other states may become
significant. This corresponds to the qualitative picture
of the adiabatic breakdown as mentioned in the last section.

\begin{figure}[h]
\includegraphics[width=3in]{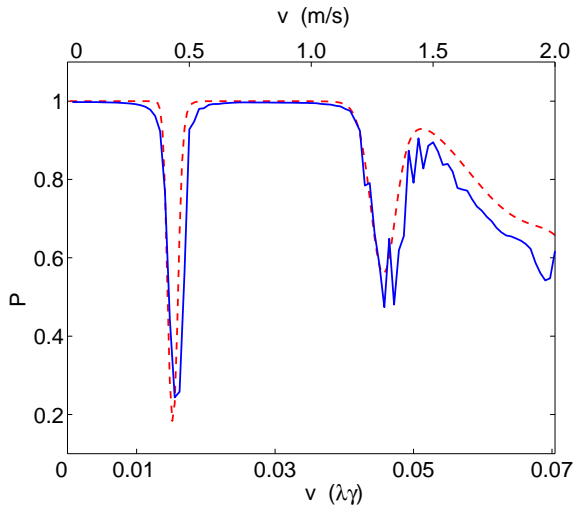}\\
\caption{The dark sate survival probability $P$ after crossing
a nodal plane when the atomic motion corresponds to
a constant speed along the cavity axis. The solid line comes
from the numerical simulation of the nodal crossing dynamics
by solving the Eqs. (\ref{eqd}) for $C_j(t)$, while the dashed
line is the prediction of our approximate analytic result
Eq. (\ref{lza}) from the Landau-Zener theory.}
\label{fig6}
\end{figure}

As a simple example, we assume the atom moves with a
constant speed $v$. At time $t=0$ when the
state transfer protocol begins,
the atom is at the peak of the cavity field where $|\sin(kz)|=1$.
In Fig. \ref{fig6} we present the results for the dark state
survival probability $P$ (as a function of $v$)
at $t=20/\gamma$, after the internal state transfer
protocol has been completed.

The two prominent features of small $P$ valleys can
be easily understood.
They correspond respectively to the crossing of a nodal plane
at instants when $h(t)$ is large as in Fig. \ref{fig5}, by
slow and fast moving atoms.
If the atomic speed is small enough,
e.g. when $v<0.01(\lambda\gamma)=0.28$ (m/s), $P$ remains essentially
unity. This is because with this speed, the atom can't
arrive at the nearest node before the state transfer protocol
is completed. When $v$ is increased to
$0.017(\lambda\gamma)=0.47$ (m/s), the first small $P$ valley shows
up, corresponding to the atom arriving at the nodal plane
at about $t=16.7/\gamma$ when $h(t)$ is significant
(near its second peak).
The second small $P$ valley corresponds to
the atomic speed of $0.049(\lambda\gamma)=1.36$ (m/s),
when the atom arrives at the nodal plane
at about $t=3.3/\gamma$ when $h(t)$ is around its first
temporal peak.
As we have analyzed in the previous section,
this second peak corresponds to the beginning of the
state transfer protocol. If the atom leaves the dark state
at this time, the whole operation will be destroyed.

We also note that the minimum of $P$
near the second small $P$ valley when $v\sim 0.049(\lambda\gamma)=1.36$ (m/s)
is larger than the minimum of the first small $P$ valley
of $v\sim 0.017(\lambda\gamma)=0.4$ (m/s).
This can be easily explained according to the analytic
result Eq. (\ref{lza}),
$P=\exp[-2\pi{h(t)}/{|kvg_0|}]$, which shows that
for the same value of $h(t)$, $P$ is larger for larger $v$.
When the atomic speed is larger than
$v\sim 0.049(\lambda\gamma)=1.36$ (m/s), the atom will cross
more than one node during the operation time,
and the dark state survival probability becomes even
smaller. We can approximate in this case $P\approx
\prod_{i=1}^nP_i$ where $n$ is the number of nodes
crossed by the atom and $P_i$ is the probability
for the atom to remain in the dark state after crossing
the $i$th node. This constitutes an excellent approximation
when $n$ is small.

\begin{figure}[h]
\includegraphics[width=3in]{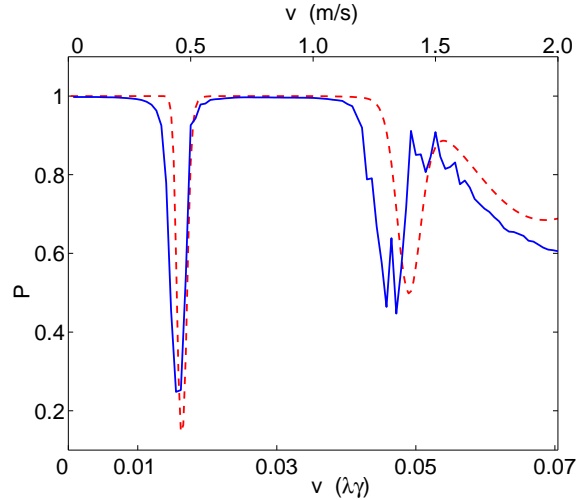}\\
\caption{The same as in Fig. \ref{fig6} but for a predetermined
oscillation of the atom as confined in an external
harmonic trap. }
\label{fig7}
\end{figure}

In practice, the atom may be trapped in an additional
potential, e.g. takes a harmonic motion instead of a straight line.
In the optimal scenario when the center of the
harmonic trap overlaps the peak of the cavity field
standing wave, and when the operation starts at the instant
when the atom is located at the trap center,
the corresponding results for this case is presented in
Fig. \ref{fig7}, where we have further assumed a typical trap
frequency $\omega_T\sim 1.32$ (MHz). We note
that the atom's final dark state survival probability
is related to its initial speed $v$ as well.

This figure resembles that of Fig. \ref{fig6}, with the
main difference being the minimum of $P$ in the
two valleys being smaller here. This is because
in an harmonic motion, its speed at
the nodal plane is smaller than its initial speed $v$.

It is remarkable that despite of the approximations used
in deriving the analytic Landau-Zener transition rate
Eq. (\ref{lza}), it gives rise to results that show
an overall agreement with the fully numerical simulations
of Eq. (\ref{eqd}). This demonstrates convincingly
that at least in the parameter regime being considered
by us, our result Eq. (\ref{lza}) captures the
complete physics involved in this model problem.

\subsection{the velocity dependence}

To gain some understanding of the effects due to the unavoidable
momentum distribution of the atom, we assume here a one dimensional
distribution (for the speed of the atomic center of mass)
\begin{eqnarray}
f(v)=\frac{1}{A}\exp\left[-\frac{(v-v_0)^2}{(\Delta v)^2}\right],
\label{fvd}
\end{eqnarray}
centered at a central velocity $v_0$ and with a distribution width $\Delta
v=1.2\times10^{-3} (\lambda\gamma)$ [$0.035$ (m/s)], or
about 10 times Cs recoil velocity. The normalization
constant is given by
\begin{eqnarray}
A(v_0,\Delta v)=\int_0^{\infty}
\exp\left[-\frac{(v-v_0)^2}{(\Delta v)^2}\right]d v.
\end{eqnarray}

In the following, we consider $v_0\in[0,5.3]\times10^{-2} (\lambda\gamma)$
i.e. $[0,1.5]$ (m/s).
We note that for each $v_0$, the above distribution (\ref{fvd}) is
essentially bounded from above by $v_{\max}\sim v_0+2\Delta v$.
When $v_0=5.3\times10^{-2} (\lambda\gamma)=1.44$ (m/s), we find
$v_{\max}\sim 5.5\times10^{-2} (\lambda\gamma)=1.5$ (m/s)
and ${v_{\max}2T_0}/({\lambda /2})=2.21$, i.e. for $v_0\leq
5.3\times10^{-2} (\lambda\gamma)$, the atom will cross at most three nodes.

The dark state survival probability as a function of $v_0$ can then be
approximately computed according to
\begin{eqnarray}
P(v_0)=\frac{B(v_0)}{A(v_0)},
\end{eqnarray}
where
\begin{eqnarray}
B(v_0)&&\approx\int_0^{\frac \lambda {4\times 2T_0}}
\exp\left[-\frac{(v-v_0)^2}{(\Delta v)^2}\right] dv \nonumber\\
&&+\int_{\frac \lambda {4\times 2T_0}}^{\frac{3\lambda }{4\times 2T_0}}\exp
\left[-\frac{(v-v_0)^2}{(\Delta v)^2}\right] P_1(v)dv  \nonumber \\
&&+\int_{\frac{3\lambda }{4\times 2T_0}}^{\frac{5\lambda }{4\times 2T_0}%
}\exp\left[-\frac{(v-v_0)^2}{(\Delta v)^2}\right]P_1(v)P_2(v)dv\nonumber
\\
&&+\int_{\frac{5\lambda}{4\times 2T_0}}^{\frac{7\lambda }{4\times 2T_0}%
}\exp \left[-\frac{(v-v_0)^2}{(\Delta v)^2}\right] P_1(v)P_2(v)P_3(v)dv,\nonumber
\end{eqnarray}
and $P_1(v)$, $P_2(v)$, and $P_3(v)$ are respectively
the dark state survival probability after
crossing the first, the second, and the third node.
In above discussion, we have assumed that the atomic initial
position is $z_0=\lambda/4$, where the cavity field has
its maximal value. The $P(v_0)$ for other values of
$z_0\neq 0$ can be obtained similarly.

\begin{figure}[h]
\includegraphics[width=3in]{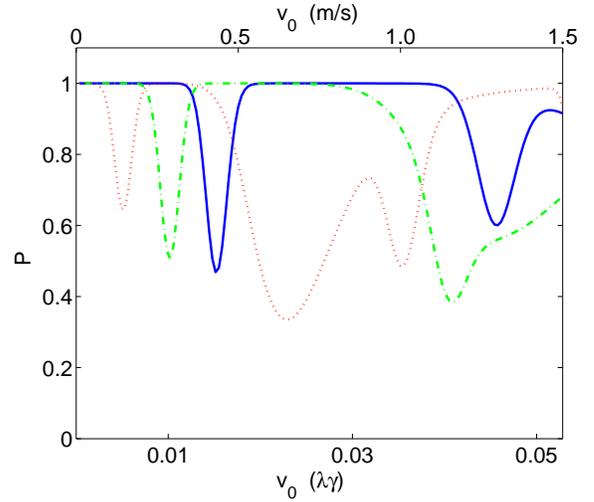}\\
\caption{$P$ as a function of $v_0$ for the initial
atomic locations along cavity axis $z_0=\lambda/4$, $\lambda/3$, and
$5\lambda/12$. }
\label{fig8}
\end{figure}

In Figure \ref{fig8}, we have presented the numerically computed
dark state survival probability $P$ as a
function of $v_0$ for several different atomic initial position
$z_0$.

\subsection{3-dimensional atomic motion}
To complete this study, we present
selective results for the 3-dimensional atomic motion
in this subsection.
We selected two different situations where the atom is initially
at the anti-nodal point of the cavity field mode,
and is taking a straight line motion that makes
an angle of $30^\circ$ or $60^\circ$ with respect to the
cavity axis. Not surprisingly, we again find excellent agreement
with our analytic Landau-Zener result Eq. (\ref{lza}),
applied appropriately as discussed earlier with the
velocity component along the cavity axis being used
to parameterize level crossing, essentially the same as the
case of the 1 dimensional model considered earlier.
\begin{figure}[h]
\includegraphics[width=3in]{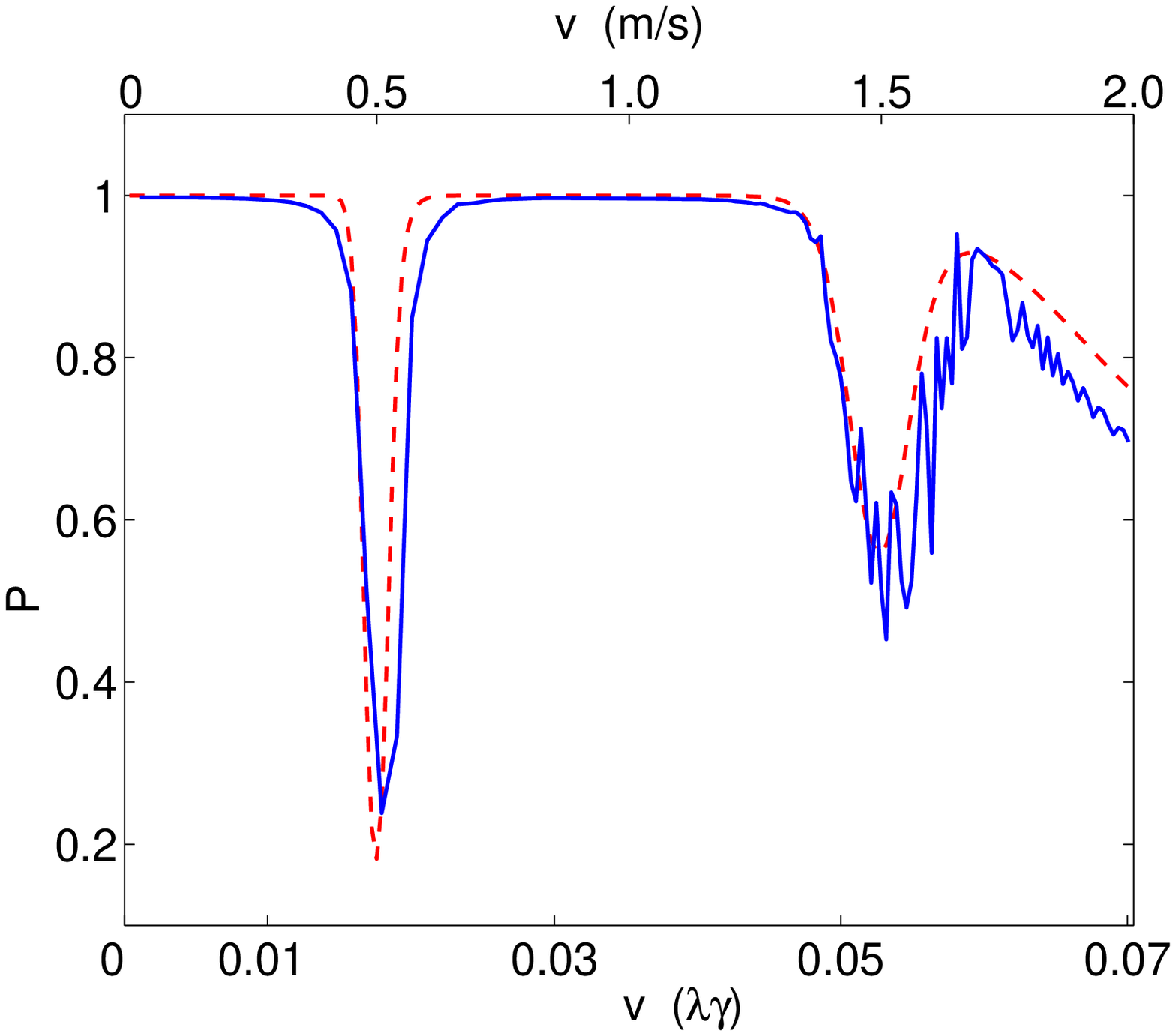}\\
\caption{The dark sate survival probability $P$ after crossing
a nodal plane when the atomic motion corresponds to
a constant speed along the direction $30^\circ$ off the
cavity axis. The solid line comes
from the numerical simulation of the nodal crossing dynamics
by solving the Eqs. (\ref{eqd}) for $C_j(t)$, while the dashed
line is the prediction of our approximate analytic result
Eq. (\ref{lza}) from the Landau-Zener theory. }
\label{fig9}
\end{figure}

\begin{figure}[h]
\includegraphics[width=3in]{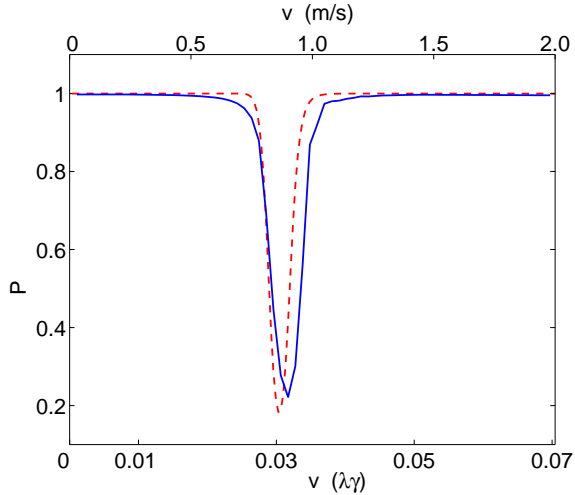}\\
\caption{The same as in Fig. \ref{fig9}
but for atomic motion $60^\circ$ off the cavity axis.
Only one small $P$ valley shows up within the
velocity range because of the large angle off the cavity axis.}
\label{fig10}
\end{figure}

\section{Summary}
In conclusion, we have studied the nonadiabatic motional effect
of a
three-level $\Lambda$-type atom Raman coupled to the standing
wave quantum field of a high Q optical cavity and an external
pump field sharing the same spatial profile.

First, making use of the Landau-Zener approximation to the
crossing of a nodal plane by the atom,
we have derived an analytic formula describing
the survival probability for the atom to stay in the so-called
motional insensitive dark state. Surprisingly, our numerical
results show that the approximation is remarkably good within
current experimental parameters, thus can be used to guide the
experimental implementation of the motional insensitive
protocol \cite{duan}.

Second, we find that the nonadiabatic motional effects is
essentially connected with the dimensionless parameter
$v\times (20/\gamma)$, the distance the atom
(with center of mass velocity $v$) travels
during the state transfer protocol of $\sim 20/\gamma$.
If this distance becomes a significant fraction of
$\lambda$, i.e. $v\times (20/\gamma)\ge \lambda/4$, or
$v\ge \lambda\gamma/(48)=0.577$ (m/s), then
nonadiabatic effect will spoil the motional
insensitive protocol in general, even if the atom is assumed
to be located initially near the antinodal planes
of $\sin(kz)=\pm 1$.

To be sure of the adiabatic following of the dark state,
one needs to assure at all times
\begin{eqnarray}
(2\pi) \frac{h(t)}{|kvg_0|}\ll 1,
\end{eqnarray}
and the number of nodes crossed is small.

\section{acknowledgement}
We thank Dr. Y. X. Miao for helpful communications.
We acknowledge the support from CNSF, the Knowledged Innovation Program
(KIP) of the Chinese Academy of Science, and the National Fundamental
Research Program of China (No. 001GB309310). L. You also acknowledges
the support of NSF.

\appendix
\section{The case of a nonzero detuning ($\Delta\neq 0$)}
\label{apda}
\begin{figure}[tbp]
\includegraphics[width=3in]{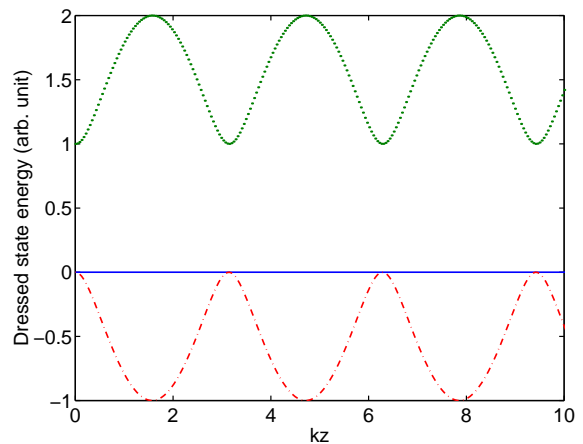}
\caption{Similar to Fig. \ref{fig2} but for $\Delta\neq 0$.}
\label{fig11}
\end{figure}

When $\Delta\neq 0$, the Hamiltonian in the interaction picture
with the same basis $\left|\{\left|e,0\right\rangle,\left|s,1\right\rangle,%
\left| g,0\right\rangle\right\}$ becomes
\begin{eqnarray}
H=\left(
\begin{array}{ccc}
\Delta & g & \Omega \\
g^{*} & 0 & 0 \\
\Omega ^{\ast } & 0 & 0
\end{array}
\right).  \label{3a}
\end{eqnarray}
The three eigenvalues are
\begin{eqnarray}
E_0 &=&0,  \nonumber \\
E_{\pm}&=&\frac{1}{2} \left(\Delta\pm\sqrt{4g^2+4\Omega^2+\Delta^2}\right),
\end{eqnarray}
with the corresponding eigenstates $\left|D\right\rangle$ and $%
\left|E_{\pm}\right\rangle$. Clearly at nodal planes when $\sin(kz)=0$, $%
\left|E_{\pm}\right|$ take their minimal values
\begin{eqnarray}
|E_{+}|_{\min} &=&\Delta,  \nonumber \\
|E_{-}|_{\min} &=&0,
\end{eqnarray}
as shown in Fig. \ref{fig11}, on inspecting of which leads to the following
two comments.

First, irrespective of whether $\Delta=0$ or $\Delta\neq 0$, the dressed
state energy levels cross at the nodal planes $\sin(kz)=0$.

Second, when $\Delta=0$, all three energy levels have the same value zero at
the nodal planes, while for $\Delta\neq 0$, only $E_{-}$ and $E_0$ take zero
values. There is a gap for $E_{+}$ whose width is $\Delta$. Thus if $\Delta $
is large enough, the transition from dark state $|D\rangle$ to $%
|E_{+}\rangle$ can be avoided, but to state $|E_-\rangle$ remains because of
the degeneracy at the crossing. The total transition probability again can be
calculated theoretically using the previously adopted Landau-Zener
approximation.

To compute the transition probability for $\Delta\neq 0$, we expand the
state of the atom plus the field in terms of the eigenbasis Eqs. (\ref{dk})
and (\ref{bk}) of the system Hamiltonian for $\Delta =0$, which is now
\begin{eqnarray}
|D\rangle &=&\frac{1}{\sqrt{1+|\alpha(t)|^2}}(|g_1,0\rangle-{\alpha(t)}
|g_0,1\rangle),  \nonumber \\
|B_{\pm}\rangle&=&\frac{1}{\sqrt{2}}(|B\rangle\pm|e,0\rangle),
\end{eqnarray}
with the corresponding eigenvalues Eq. (\ref{edress}) reexpressed in this
appendix as
\begin{eqnarray}
\epsilon_0 &=&0,  \nonumber \\
\epsilon_{\pm} &=&\pm\chi[r(t)]|g_0|\sqrt{1+|\alpha(t)|^2}.
\end{eqnarray}

As before in Eq. (\ref{bk}), we have introduced
\begin{eqnarray}
|B\rangle =\frac{1}{\sqrt{1+|\alpha(t)|^2}}
[|g_1,1\rangle+\alpha(t)|g_0,0\rangle],
\end{eqnarray}
and the parameter $r_0$ has been assumed to be unity.
Then the system Hamiltonian including $\Delta$ is
\begin{eqnarray}
H=\Delta |e,0\rangle\!\langle e,0|
+\epsilon_{+}|B_{+}\rangle\!\langle B_{+}|
+\epsilon_{-}|B_{-}\rangle\!\langle B_{-}|.
\label{hd}
\end{eqnarray}
Noting that
\begin{eqnarray}
|e,0\rangle =\frac{1}{\sqrt{2}}(|B_{+}\rangle-|B_{-}\rangle),
\end{eqnarray}
we rewrite Eq. (\ref{hd}) as
\begin{eqnarray}
H &=&\left(\epsilon_{+}+\frac{\Delta}{2}\right)|B_{+}\rangle \!\langle
B_{+}|+\left(\epsilon_{-}+\frac{\Delta}{2}\right)
|B_{-}\rangle\!\langle B_{-}|  \nonumber \\
&&-\frac{\Delta}{2}(|B_{+}\rangle\!\langle B_{-}|
+|B_{-}\rangle\!\langle B_{+}|).
\end{eqnarray}

Expanding the quantum state as in Eq. (\ref{eps})
\begin{eqnarray}
|\Psi(t)\rangle &=&C_0|D\rangle
+C_{+}e^{-i\int_0^t[\epsilon_{+}(t')
+\Delta ]dt'}|B_{+}\rangle  \nonumber \\
&&+C_{-}e^{-i\int_0^t[\epsilon_{-}(t')+\Delta ]dt'}|B_{-}\rangle,
\end{eqnarray}
we obtain the following equations
\begin{eqnarray}
\dot{C}_0 &=&-(C_{+}
e^{-i\int_0^t[\epsilon (t')+\Delta ]dt'}
+C_{-}e^{-i\int_0^t[-\epsilon (t')+\Delta ]dt'})K, \nonumber\\
\dot{C}_{+} &=&C_0e^{i\int_0^t[\epsilon (t')+\Delta ]dt'}K\nonumber\\
&&+i\frac \Delta 2(C_{-}e^{i\int_0^t2\epsilon (t')dt'}+C_{+}), \nonumber\\
\dot{C}_{-} &=&C_0e^{i\int_0^t[-\epsilon (t')+\Delta ]dt'}K
\nonumber\\
&&+i\frac \Delta 2(C_{+}e^{-i\int_0^t2\epsilon (t')dt'}+C_{-}).
\label{deq}
\end{eqnarray}
Since
\begin{eqnarray}
\dot{C}_{\pm}=e^{\pm i\int_0^t2\epsilon (t')dt'}\dot{C}_{\mp},
\end{eqnarray}
we can integrate it to obtain
\begin{eqnarray}
C_{\pm} &=&\int\dot{C}_{\mp}(t')e^{\pm i\int_0^{t'}2\epsilon
(t'')dt''}dt' \nonumber\\
&\simeq &C_{\mp}e^{\pm i\int_0^t2\epsilon (t')dt'},
\end{eqnarray}
where we have neglected the ``small term"
$\int C_{\mp}(t')d\exp[\pm i\int_0^{t'}2\epsilon (t'')dt'']$
(as in Sect. \ref{sect3} and appendix \ref{apdc}).
Then according to Eq. (\ref{deq}), we have
\begin{eqnarray}
\dot{C}_{+} &=&C_0e^{i\int_0^t[\epsilon (t')+\Delta ]dt'}K+i\Delta C_{+}, \nonumber \\
\dot{C}_{-} &=&C_0e^{i\int_0^t[-\epsilon (t')+\Delta
]dt'}K+i\Delta C_{-}.
\end{eqnarray}

Assuming $C_{\pm}=e^{i\Delta t}\xi_{\pm}$,
the above equation can be expressed as
\begin{eqnarray}
\dot{\xi}_{\pm}=C_0e^{\pm i\int_0^t\epsilon (t')dt'}K,
\end{eqnarray}
which when coupled with the equation
\begin{eqnarray}
\dot{C}_0 &=&-(C_{+}e^{-i\int_0^t[\epsilon(t')+\Delta]dt'}
+C_{-}e^{-i\int_0^t[-\epsilon(t')+\Delta]dt'})K  \nonumber \\
&=&-(\xi_{+}e^{-i\int_0^t\epsilon(t')dt'}
+\xi_{-}e^{-i\int_0^t\epsilon(t')dt'})K,
\end{eqnarray}
is formally the same as equations for $\Delta =0$.
Thus we obtain the same result
\begin{eqnarray}
|C_0|^2 &=&1-|\xi_{+}|^2-|\xi_{-}|^2  \nonumber \\
&=&\exp\left[-2\pi \frac{h(t)}{|kvg_0|}\right].
\end{eqnarray}

\section{The Linear Approximation}
\label{apdb}
Within the discussion as in subsection \ref{subsecb}, we mapped
our level crossing problem into the well known problem of Landau-Zener
transition.

As was shown before, the adiabatic condition is
\begin{eqnarray}
\left|\frac{f(t)}{\sin(kz)}\right|\ll 1,
\end{eqnarray}
with the typical behavior for $f(t)$ as shown in Fig. \ref{fig4}. Taking $%
f(t)\leq 0.035$, we see that within the domain of $|\sin(kz)|<0.7$, we have
\begin{eqnarray}
\min\left|\frac{f(t)}{\sin(kz)}\right| <0.05,
\end{eqnarray}
although still much less than 1. Thus we can define the domain $%
|\sin(kz)|<0.7$ as the domain of validity where the adiabatic condition is
marginal. In this domain, the error of the linear approximation $\sin(kz)\sim
kz$ is about $10\%$.

\section{the small term}
\label{apdc}
In this appendix, we provide the justification for the neglect of
the second term of Eq. (\ref{sterm}).

Given that initially the atom is in the dark state, we need $%
C_{+}(t')$ to be small in order to maintain adiabatic operation.
Thus, we approximate
\begin{eqnarray}
C_{-} &=&e^{-2i\int_0^{t'}\varepsilon
(t^{\prime\prime})dt^{\prime\prime}/\hbar}C_{+}  \nonumber \\
&&+2i\int_{-u}^te^{-2i\int_0^{t'}\varepsilon
(t^{\prime\prime})dt^{\prime\prime}/\hbar} {\varepsilon (t')}%
C_{+}(t')dt'/{\hbar},  \label{ll}
\end{eqnarray}
where $u$ is the end of the time domain and assumed to satisfy $%
|\sin(kvu)|=0.7$.

We first approximate the second term of Eq. (\ref{ll}) according to
\begin{eqnarray}
&&\left|2i\int_{-u}^te^{-2i\int_0^{t'}\varepsilon
(t^{\prime\prime})dt^{\prime\prime}/\hbar}{\varepsilon(t')}%
C_{+}(t')dt'/{\hbar}\right|  \nonumber \\
&&\sim \left|2i\int_{-u}^te^{-2i\int_0^{t'}\varepsilon
(t^{\prime\prime})dt^{\prime\prime}/\hbar}{\varepsilon(t')}%
dt'C_{+}(t)/{\hbar}\right|.
\end{eqnarray}
We note that
\begin{eqnarray}
S &=&\left| 2i\int_{-u}^te^{-2i\int_0^{t'}\varepsilon
(t^{\prime\prime})dt^{\prime\prime}/\hbar}{\varepsilon(t')}%
dt'/{\hbar}\right|  \nonumber \\
&=&2\left|\int_{-u}^te^{-2i\int_0^{t'}kvt^{\prime\prime}g_0
\sqrt{1+|\alpha|^2}dt^{\prime\prime}}kvt'g_0\sqrt{1+|\alpha|^2}%
dt'\right|  \nonumber \\
&\approx &2\left|\int_{-u}^te^{-ikvt^{'2}g_0\sqrt{1+|\alpha|^2}%
}kvt'g_0\sqrt{1+|\alpha|^2}dt'\right|  \nonumber \\
&=&2\left|\int_{-\Omega u}^{\Omega
t}e^{-i\tau^{\prime2}}\tau'd\tau'\right| ,
\end{eqnarray}
where we have approximated $\alpha(t)$ as a constant and denoted
\begin{eqnarray}
\Omega &=&\sqrt{kvg_0\sqrt{1+|\alpha|^2}},  \nonumber \\
\tau'&=&\Omega t'.
\end{eqnarray}
Thus
\begin{eqnarray}
S &=&2\left|\int_{-\Omega u}^{\Omega
t}e^{-i\tau^{\prime2}}\tau'd\tau'\right| \\
&=&2\left|\int_{-\sqrt{kvg_0\sqrt{1+|\alpha|^2}}\frac{0.77}{kv}}^{\sqrt{%
kvg_0\sqrt{1+|\alpha|^2}}t}e^{-i\tau^{\prime2}}\tau'd\tau'%
\right|
\end{eqnarray}
where we have used $\sin^{-1}0.7=0.77$.

We take the worst case and use the value of $\alpha $ when $\dot{\alpha}$
is maximum. This leads to
\begin{eqnarray}
\alpha &=&30\times \exp \left[ -\frac{\left( 1\times 10^{-7}-6\times
0.5\times 10^{-7}\right) ^2}{10^{-14}}\right]  \nonumber \\
&\approx &0.55,  \nonumber \\
\sqrt{1+|\alpha|^2} &=&1.14,
\end{eqnarray}
and $\sqrt{kvg_0\sqrt{1+|\alpha|^2}}=5.0\times 10^7\sqrt{v}$. If we now
take $t={s}/{(kv)}$, we find
\begin{eqnarray}
S &=&2\left|\int_{-\sqrt{kvg_0\sqrt{1+|\alpha|^2}}\,\frac{0.77}{kv}}^{%
\sqrt{kvg_0\sqrt{1+|\alpha|^2}}\frac s{kv}}e^{-i\tau ^{\prime 2}}\tau'
d\tau'\right| \\
&=&2\left|\int_{-\frac{5.69}{\sqrt{v}}}^xe^{-i\tau^{'2}}\tau'd\tau'\right|,
\end{eqnarray}
with
\begin{eqnarray}
x=\sqrt{kvg_0\sqrt{1+|\alpha|^2}}\,\frac s{kv}.
\end{eqnarray}

The oscillating behaviors of $S$ for
$v=3.5\times 10^{-2}(\lambda\gamma)$ [$1$ (m/s)],
$1.2\times 10^{-2}(\lambda\gamma)$ [$0.35$ (m/s)],
and $v=3.5\times 10^{-3}(\lambda\gamma)$ [$0.1$ (m/s)] are
shown below in Figs. \ref{fig12}, \ref{fig13}, and \ref{fig14}.
\begin{figure}[h]
\includegraphics[width=3in]{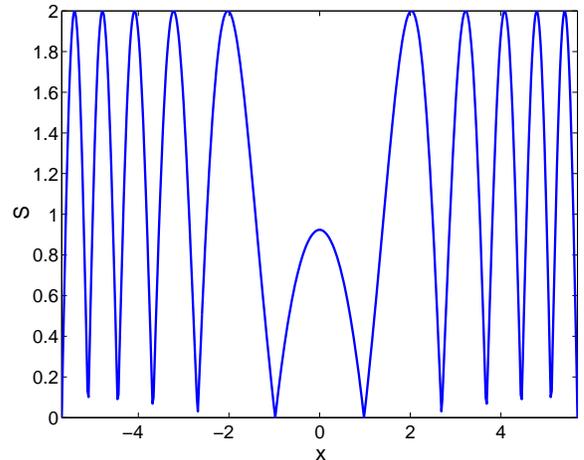}
\caption{The term $S(x)$ as a function of $x$ for
$v=3.5\times 10^{-2} (\lambda\gamma)$.}
\label{fig12}
\end{figure}

\begin{figure}[h]
\includegraphics[width=3in]{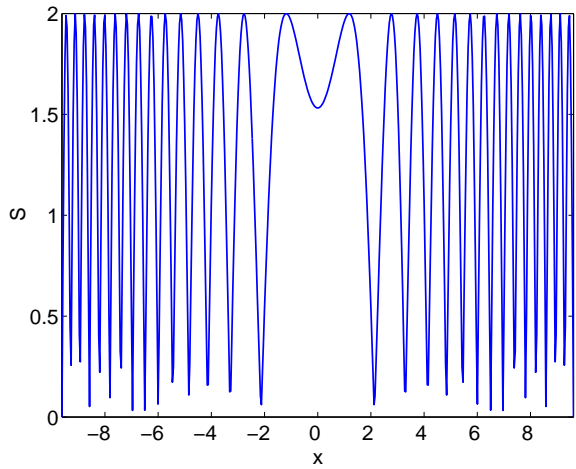}
\caption{The same as in Fig. \ref{fig12} but for
$v=1.2\times 10^{-2} (\lambda\gamma)$.}
\label{fig13}
\end{figure}

\begin{figure}[h]
\includegraphics[width=3in]{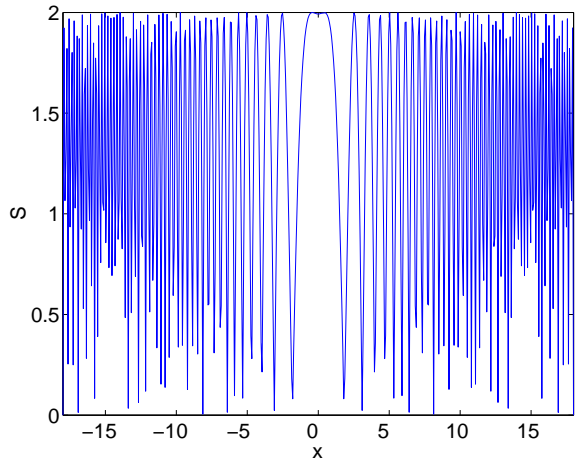}
\caption{The same as in Fig. \ref{fig12} but for
$v=3.5\times 10^{-3} (\lambda\gamma)$.}
\label{fig14}
\end{figure}

We see that the amplitude of $S$ is about 2, not really a small value. On
the other hand, $S$ is a rapid oscillation function of time $t$, thus does
not lead to much effect during the dynamic evolution. We believe this is the
reason why our Landau-Zener result based on the neglect of this ``small
term" is justified by the numerical simulations.

\end{document}